# Comparison of kinetic and dynamical models

# of DNA-protein interaction and facilitated diffusion

Ana-Maria FLORESCU and Marc JOYEUX [(#)]

*Laboratoire de Spectrométrie Physique (CNRS UMR 5588),*

*Université Joseph Fourier - Grenoble 1,*

*BP 87,*

*38402 St Martin d'Hères*

*France*

[(#)] email : Marc.Joyeux@ujf-grenoble.fr

**Abstract :** It has long been asserted that proteins like transcription factors may locate their target in DNA sequences at rates that surpass by several orders of magnitude the three-dimensional diffusion limit thank to facilitated diffusion, that is the combination of one-dimensional (sliding along the DNA) and three-dimensional diffusion. This claim has been supported along the years by several mass action kinetic models, while the dynamical model we proposed recently (J. Chem. Phys. 130, 015103 (2009)) suggests that acceleration of targeting due to facilitated diffusion cannot be large. In order to solve this apparent contradiction, we performed additional simulations to compare the results obtained with our model to those obtained with the kinetic model of Klenin *et al* (Phys. Rev. Letters 96, 018104 (2006)). We show in this paper that the two models actually support each other and agree in predicting a low efficiency for facilitated diffusion. Extrapolation of these results to real systems even indicates that facilitated diffusion necessarily slows down the targeting process compared to three-dimensional diffusion.

# 1. Introduction

Prof. S.E. Halford (University of Bristol, UK) recently published an article entitled *"An end to 40 years of mistakes in DNA-protein association kinetics ?"* [1]. In this article, he fights against the popular belief that some proteins, like transcription factors, can bind to their specific DNA targets at rates that surpass the diffusion limit, that is the rate at which the protein and the DNA collide with each other as a result of 3-dimensional (3D) thermal diffusion. According to him, there is *"no known example of a protein binding to a specific DNA site at a rate above the diffusion limit"* [1].

The whole story actually started with the pioneering work of Riggs, Bourgeois and Cohn [2], which was completed shortly after the first demonstration that proteins can bind to specific DNA sequences [3,4]. Riggs *et al* reported that, in a buffer containing KCl, Tris-HCl and magnesium acetate at 0.01 M concentrations (M means mole per liter), the Lac repressor binds to its operator site at a rate of about $7 \times 10^9$ $M^{-1}$ $s^{-1}$ [5]. This value is one to two orders of magnitude larger than the one that is generally assumed for the diffusion limit of protein-DNA association, that is $10^8$ $M^{-1}$ $s^{-1}$ [5,6].

Elaborating on the early remarks of Adam and Delbrück [7] and Richter and Eigen [8], that reducing the dimensionality of diffusion-based reactions can greatly increase their efficiency, Berg and coworkers concluded that the result of Riggs *et al* proves that DNA-protein interactions cannot rely on pure 3D diffusion, but must instead involve alternate sequences of 3D diffusion of the protein in the buffer and 1-dimensional (1D) sliding of the protein along the DNA sequence. This process is known as *"facilitated diffusion"*. Most mass action kinetic models developed today to investigate facilitated diffusion are still based on the model that Berg and coworkers proposed to legitimate their claim [9-12]. Quite interestingly, several recent single molecule experiments confirmed that the investigated proteins indeed find their specific targets thanks to a mixing of 1D and 3D diffusion [13-21]. The facilitated diffusion mechanism proposed by Berg and coworkers therefore seems to be indisputable.

The assertion that facilitated diffusion may greatly enhance DNA-protein association rates - or equivalently that the rate measured by Riggs *et al* is much larger than the diffusion limit - is instead more questionable. It was precisely the purpose of Ref. [1] to highlight this point. Estimation of the diffusion limit is based on Smoluchowski's rate constant for a reaction limited by the diffusional collision frequency, which writes [22]

$$k_{Smol} = 4\pi (1000 N_{Avog})(D_A + D_B)(r_A + r_B) , \qquad (1.1)$$

in units of $M^{-1} s^{-1}$. $N_{Avog}$ is Avogadro's number, $D_A$ and $D_B$ are the 3D diffusion constants of the colliding species A and B (in units of $m^2 s^{-1}$), and $r_A$ and $r_B$ their reaction radii (in meter units). Riggs *et al* pointed out that the diffusion constant of DNA is negligible compared to that of the protein and consequently estimated that $D_A + D_B \approx 0.50 \times 10^{-10}$ $m^2 s^{-1}$ on the basis of the 150000 molecular weight of the Lac repressor (this is very close to the value obtained from Einstein's formula for the diffusion constant of a sphere). They further considered that the reaction radius $r_A + r_B$ is of the order of 0.5 nm, that is approximately the size of a base or an amino-acid. By plugging these numerical values in Eq. (1.1), one obtains $k_{Smol} \approx 2 \times 10^8$ $M^{-1} s^{-1}$, that is about 35 times less than the measured value.

However, the crucial point is that Eq. (1.1) is valid only if molecules A and B have no net charge or if these charges are neutralized by counterions [23]. If this is not the case, then the association rate for free diffusion must be modified to include an electrostatic factor $f_{elec}$ [1,9]

$$k = k_{Smol} f_{elec} . \qquad (1.2)$$

$f_{elec}$ is larger than 1 if the interacting surfaces of A and B possess opposite charges. It is instead comprised between 0 and 1 if the sign of the charges is the same. Moreover, $f_{elec}$ usually tends towards 1 when ionic strength is increased, because there are more and more counterions to neutralize the charges on the interacting surfaces of A and B. The rate of $7 \times 10^9$ M$^{-1}$ s$^{-1}$ reported by Riggs *et al* [2] was precisely measured at very low ionic strength. Moreover, Riggs *et al* also reported that they measured an association rate about 100 times smaller (and consequently close to the diffusion limit) in a buffer containing a concentration of KCl of 0.1 M instead of 0.01 M. They concluded that "this sensitivity to ionic strength strongly suggests that the binding of repressor to operator is aided by electrostatic attraction between the negatively charged DNA chain and positively charged groups of the repressor", because "such long range attractive forces between repressor and DNA would be expected to accelerate greatly the association reaction over that predicted by Smoluchowski's equation" [2]. Clearly, Riggs *et al* did not see any contradiction between the rate measured at 0.01 M salt concentration and Smoluchowski's one, a point which seems to have been largely overlooked for decades [1].

At that point, it should be mentioned that all subsequent experiments performed with the Lac operator consistently reported rates of the order of $10^{10}$ M$^{-1}$ s$^{-1}$ in the absence of salt, which however fell almost logarithmically down to $10^8$ M$^{-1}$ s$^{-1}$ as salt concentration was increased [10,24,25]. Stated in other words, for the Lac repressor/operator system, $f_{elec}$ is very large, presumably close to or larger than 100, in the absence of salt but decreases down to 1 at physiological concentrations (see Appendix A for a more detailed numerical estimation). Last but not least, it appears that all of the (few) other examples reported in the literature of proteins binding to their targets at rates significantly above $10^8$ M$^{-1}$ s$^{-1}$ were investigated at low salt concentrations (see [1] and references therein).

We recently proposed a "dynamical" (or "molecular mechanical") model for non-specific DNA-protein interactions and facilitated diffusion, that is a microscopic model which relies uniquely on the definition of a Hamiltonian describing all interactions inside the model cell and the choice of equations of the motion [26,27]. This model is based on the "worm-like chain and beads" or "bead-spring" model for DNA, which was developed by several groups [28-33]. We added to this DNA model one or several interconnected beads to describe the protein and paid special attention to the terms of the Hamiltonian that model the interactions between DNA and the protein [26,27]. We used the Brownian dynamics algorithm of Ermak and McCammon [34], which includes hydrodynamic interactions, to propagate classical trajectories. This confirmed that our model satisfactorily reproduces the succession of motions by which the protein samples the DNA sequence, that is 3D diffusion in the buffer, 1D sliding along the sequence, short hops between neighboring sites, long hops between more distant sites, and intersegmental transfers. We also showed that the portion of time during which the protein slides along the DNA can be varied from nearly 0 to 1 by increasing the charge(s) placed at the center of the protein bead(s), which provides a straightforward method to investigate quite different dynamical regimes. One of the main results of [26,27] is that this model suggests that, whatever the portion of time the protein spends in 1D sliding along DNA, the facilitated diffusion mechanism cannot increase the sampling speed by more than 70% when the protein is described by a single bead [26]. The maximum speed increase may even be much smaller – and vanish – when the protein is modelled as a set of interconnected beads [27]. Results obtained with this model therefore agree with the claim of Prof. Halford [1].

It seems, however, that the results obtained with our dynamical model differ from those obtained with mass action kinetic models, that is models based on an *a priori* scenario of site targeting and the definition of a more or less extended list of phenomenological rates and probabilities. These later models indeed usually predict that the facilitated diffusion mechanism is capable of increasing the targeting speed by one to several orders of magnitude [35-39]. We therefore tried to compare our model to kinetic ones, in order to understand what this difference is due to. It rapidly appeared that several of the kinetic models more or less explicitly disregard exact numerical coefficients. For example, [35,36] neglect the $4\pi$ coefficient in Eq. (1.1). As a result, these models provide simple analytical results that are very useful to decipher the *scaling* dependence of the targeting speed on major system parameters, but they cannot provide reliable estimates of the speed increase due to facilitated diffusion. Other models rely heavily on terms or expressions, which are quite difficult to relate to the properties of real molecules and/or quantities derived from dynamical systems (see for example the last term in the right-hand side of Eq. (6) of [39]). We essentially found only one kinetic model to which to compare the results obtained with our dynamical model. Using older calculations of Szabo *et al* [40], Klenin, Merlitz, Langowski and Wu indeed obtained that the mean time of the first arrival of the protein at the target of radius *a* can be written in the form [38]

$$\tau = (\frac{V}{8D_{3D}\xi} + \frac{\pi L \xi}{4D_{1D}})[1 - \frac{2}{\pi}\arctan(\frac{a}{\xi})] \ , \tag{1.3}$$

where $D_{1D}$ and $D_{3D}$ are the diffusion coefficients of the protein in the buffer and along the DNA sequence, respectively, $V$ is the volume of the buffer, $L$ the length of the DNA sequence, and $\xi$ the distance

$$\xi = \sqrt{\frac{1}{2\pi}\frac{V}{L}\frac{D_{1D}}{D_{3D}}\frac{\tau_{1D}}{\tau_{3D}}} \tag{1.4}$$

($\tau_{1D}$ and $\tau_{3D}$ are the average times the protein spends in the bound and free states, respectively). The accuracy of Eq. (1.3) was checked for the simple system where the protein is described as a random walker that is allowed to enter freely in the neighbourhood of the DNA but has a given finite probability to exit this volume at each time step [38,41,42].

As will be shown below, all the quantities that appear in Eqs. (1.3) and (1.4) can also be derived from Brownian dynamics simulations, and the mean time of first arrival $\tau$ can furthermore be related to the rate constant $k$. The purpose of this paper is consequently to compare the results obtained with our dynamical model to those obtained with the kinetic model of Klenin *et al*, and to see what conclusions regarding facilitated diffusion in real systems can be drawn from this comparison. The remainder of the article is organized as follows. The dynamical model and Brownian dynamics simulations are briefly described in Sect. 2 for the sake of self-consistency. The relation between the mean time of first arrival $\tau$ and the rate constant $\kappa$, which we use to compare the dynamical and kinetic models, is derived in Sect. 3. The results of Brownian dynamics simulations and the extraction of the physical parameters that appear in the kinetic model are then presented in Sect. 4. These results and the agreement between the dynamical and kinetic models are discussed in Sect. 5. We finally discuss in Sect. 6 the conclusions regarding facilitated diffusion in real systems, which can be drawn from this comparison.

## 2. Dynamical model and Brownian dynamics simulations

As illustrated in Fig. 1, our model consists of DNA and a protein enclosed in a sphere, which describes the cell or its nucleus [26,27]. The protein is modeled by a single bead of hydrodynamic radius $a_{\text{prot}} = 3.5$ nm with an effective charge $e_{\text{prot}}$ placed at its center. As discussed in [26], DNA is not modeled as a single long chain, in order to avoid excessive DNA curvature at the cell walls. It is instead modeled as a set of $m$ disconnected smaller chains (hereafter called *segments*), each one consisting of $n$ beads that are separated at equilibrium by a distance $l_0 = 5.0$ nm. Each bead, which represents 15 base pairs, has a hydrodynamic radius $a_{DNA} = 1.78$ nm and an effective charge $e_{DNA} = -0.243 \times 10^{10} l_0 \bar{e} \approx -12 \bar{e}$ placed at its center ($\bar{e}$ is the absolute charge of the electron). In [26,27], the radius $R_0$ of the sphere was chosen so that the density of bases inside the cell is close to the physiological value. In the present work, we instead varied the DNA concentration, in order to better check the respective influences of 1D and 3D motions (see below). As pointed out in [35], the volume $V$ of the buffer is connected to the total DNA length $L$ according to $V = w^2 L$, where $w$ represents roughly the spacing of nearby DNA segments ($w$ is of the order of 30-50 nm for both prokaryote and eukaryote cells). $R_0$, $m$ and $n$ therefore fulfill the relation $\frac{4}{3}\pi R_0^3 = w^2 m n l_0$. Moreover, we chose the length of each DNA segment to be approximately equal to the radius of the cell, that is $n l_0 \approx R_0$, so that (i) the cell is rather homogeneously filled with DNA, (ii) end effects are negligible, and (iii) excessive curvature of DNA segments touching the cell wall is avoided. Practically, we kept the total number of DNA beads constant ($m \times n = 4000$) and investigated the dynamics of the following systems :

(a) $w$=18 nm, $m$=160, $n$=25, and $R_0$=0.116 μm

(b) $w$=32 nm, $m$=125, $n$=32, and $R_0$=0.170 μm

(c) $w$=45 nm, $m$=80, $n$=50, and $R_0$=0.213 μm

(d) $w$=135 nm, $m$=40, $n$=100, and $R_0$=0.443 μm.

The potential energy $E_{pot}$ of the system consists of three terms

$$E_{pot} = V_{DNA} + V_{DNA/prot} + V_{wall} ,  \qquad (2.1)$$

where $V_{DNA}$ describes the potential energy of the DNA segments and the interactions between them, $V_{DNA/prot}$ stands for the interactions between the protein bead and DNA segments, and $V_{wall}$ models the interactions with the cell wall, which maintain the protein bead and the DNA segments inside the cell. The expressions for $V_{DNA}$, $V_{wall}$ and $V_{DNA/prot}$ are given in Eqs. (2.2), (2.4) and (2.6) of [26], respectively. It is important to realize that the interaction energy between a DNA bead and the protein one must be minimum for a distance close to the sum of their radii, $\sigma = a_{DNA} + a_{prot} = 5.28$ nm, in order for 1D sliding to take place. The expression for the excluded volume term $E_{ev}$ in eq (2.6) of [26] insures that this is indeed the case and that the position of the minimum does not depend on the protein charge $e_{prot}$. As will be discussed in some detail below, we also considered the case where the DNA-protein interaction is purely repulsive (see Fig. 1 of [26]). The repulsive potential is obtained by keeping only the repulsive part of the interaction potential for $e_{prot}/e_{DNA} = 0.3$, that is by shifting the energies so that the minimum energy is zero and setting to zero the interaction energy for DNA-protein distances larger than the position of the minimum.

We performed two different sets of simulations. For the first set, the effect of the buffer was modeled in the simple standard way by introducing a dissipative and a stochastic term in the equations of motion. For the second set of simulations, we additionally took hydrodynamic interactions (HI) into account, that is the influence of the motion of a given bead on the motion of the neighboring ones. For simulations where HI were neglected, the positions of all of the beads were updated at each time step according to Eq. (2.10) of [26], which is just the discretization of standard Langevin equations without inertial contributions. For simulations which took HI into account, the positions of the protein bead and the 100 DNA beads closest to it were instead updated according to the BD algorithm of Ermak and McCammon [34], that is Eq. (2.8) of [26]. Since we are essentially interested in the time evolution of the number of different DNA beads visited by the protein, the results obtained with this procedure are identical to those obtained by using the algorithm of Ermak and McCammon to update the positions of all of the 4000 DNA beads, while the required CPU time is orders of magnitude times shorter [26].

For the sake of completeness, let us finally note that all simulations were performed with a time step $\Delta t = 400$ ps and began with a 40 µs thermalization cycle. Moreover, all the results presented below correspond to the average of at least four simulations with different random initial conditions.

## 3. Method of comparison

In this section, we describe the method we used to compare our model [26] with that of Klenin *et al* [38].

The basic information, which we extract from Brownian dynamics simulations, is the time evolution of the number $N(t)$ of different DNA beads visited by the protein. As will be illustrated and discussed in more detail below, we found that, for all the simulations we ran, $N(t)$ follows very precisely the same law as we already observed in our previous studies [26,27], that is

$$\frac{N(t)}{mn} = 1 - \exp\left(-\kappa \frac{t}{mn}\right), \tag{3.1}$$

where $mn = 4000$ is the total number of DNA beads. By inverting this relation, one obtains that the time $t_k$ of first arrival at the *k*th distinct bead is

$$t_k = -\frac{mn}{\kappa} \ln\left(1 - \frac{k}{mn}\right). \tag{3.2}$$

This relation is, however, necessarily wrong for the last DNA bead ($k = mn$), since it predicts that it takes an infinite time for the protein to reach this bead, while this time must be finite. By computing the mean time of first arrival $\tau$ over the other $mn-1$ beads, one obtains

$$\tau = \frac{1}{mn-1} \sum_{k=1}^{mn-1} t_k = -\frac{1}{\kappa} \sum_{k=1}^{mn-1} \ln\left(1 - \frac{k}{mn}\right), \qquad (3.3)$$

which, for large values of $mn$, is very close to

$$\tau \approx \frac{mn}{\kappa}. \qquad (3.4)$$

It can be checked numerically that the validity of Eq. (3.4) degrades only slowly when the average in Eq. (3.3) is calculated over $mn-10$ or $mn-100$ beads instead of $mn-1$. This indicates that the validity of Eq. (3.4) does not depend too sensitively on the exact asymptotic behaviour of $N(t)$ close to $mn$ (note that we checked that Eq. (3.1) remains valid even when the protein has already visited more than 99.5% of the total number of DNA beads). Moreover, it is also possible to check that, for a pure diffusive 3D motion, the rate $\kappa$ obtained from the time evolution of $N(t)$ and the mean time of first arrival $\tau$ obtained from Klenin *et al*'s formula in Eq. (1.3) are indeed related through Eq. (3.4). For that purpose, let us first note that, as long as $N(t)$ remains small compared to $mn$, expansion up to first order of the exponential in Eq. (3.1) leads to

$$N(t) \approx \kappa t. \qquad (3.5)$$

This means that the number of different DNA beads visited by the protein increases linearly with time as long as saturation does not set in. At first, we believed that this indicates that the global 3D motion of the protein is not diffusive [26]. However, we then realized that, according to a rather old mathematical result known as the "volume of the Wiener sausage" [43], the volume $V(t)$ visited by a diffusive solid with diffusion coefficient $D_{3D}$ increases linearly with time according to [44,45]

$$V(t) = 4\pi \, \delta \, D_{3D} \, t, \qquad (3.6)$$

where $\delta$ is the maximum distance away from the central Brownian motion of each point of the solid. If one assumes that DNA is homogeneously distributed in the cell and that the diffusive motion of DNA is slow compared to that of the protein, then $N(t)$ and $V(t)$ may be related through $N(t) = c\,V(t)$, where $c$ is the concentration of DNA beads. Combination of Eqs. (3.5) and (3.6) therefore leads to

$$\kappa = 4\pi D_{3D} \delta c \ . \tag{3.7}$$

Stated in other words, if one assumes that $\delta = \sigma = a_{DNA} + a_{prot}$, then $\kappa$ is just Smoluchowski's rate (Eq. (1.1)) converted to $s^{-1}$ units. As will be discussed below, simulations performed with the repulsive DNA/protein interaction potential and without HI agree very well with Eq. (3.7).

Moreover, one easily checks that, in the absence of 1D sliding ($\tau_{1D} \to 0$) and for a radius $a$ equal to $\delta$, the mean time of first arrival obtained from Klenin *et al*'s relation in Eq. (1.3) tends towards

$$\tau = \frac{V}{4\pi D_{3D} \delta} = \frac{mn}{4\pi D_{3D} \delta c} \ . \tag{3.8}$$

Comparison of Eqs. (3.7) and (3.8) confirms that $\kappa$ and $\tau$ are indeed related through Eq. (3.4) for 3D diffusion.

The strategy we adopted to compare our model with that of Klenin *et al* therefore consists in extracting several quantities from the simulations we ran. On one side, we directly estimated the rate constant $\kappa$ from each simulation by fitting the computed evolution of $N(t)$ against Eq. (3.1). On the other side, we also derived numerical values for $D_{1D}$, $D_{3D}$, $\tau_{1D}$ and $\tau_{3D}$ from the same simulations (see below for more detail). We used these values to compute the mean time of first arrival $\tau$ according to Klenin *et al*'s formula in Eq. (1.3). We finally converted $\tau$ to a rate constant $\kappa$ by using Eq. (3.4) and compared it to the value of $\kappa$ deduced from the time evolution of $N(t)$.

## 4. Results

In this section, we report the values of the rate constant $\kappa$, the fraction of time $\rho_{1D}$ that the protein spends sliding along the DNA, and the diffusion coefficients $D_{1D}$ and $D_{3D}$, which we extracted from Brownian dynamics simulations performed with our model under many different conditions. We then use Klenin *et al*'s formula in Eq. (1.3), in conjunction with Eq. (3.4), to get a second estimate of the rate constant $\kappa$ from the kinetic model.

However, let us first emphasize that all estimations reported in this section require the definition of a threshold distance between the position $\mathbf{r}_{j,k}$ of bead $k$ of DNA segment $j$ and the position $\mathbf{r}_{prot}$ of the protein, such that the two beads are assumed to be interacting whenever the distance between them is smaller than this threshold. In what follows, we systematically considered two different thresholds. More precisely, we assumed that the DNA and protein beads are interacting either if $\|\mathbf{r}_{j,k} - \mathbf{r}_{prot}\| \leq \sigma$ or if $\|\mathbf{r}_{j,k} - \mathbf{r}_{prot}\| \leq 1.5\sigma$, where $\sigma = a_{DNA} + a_{prot} = 5.28$ nm is the sum of the radii of the DNA and protein beads. We will see later that the choice of the threshold is rather important for the purpose of comparing kinetic and dynamical models.

### A - *Estimation of the rate constant $\kappa$ from the time evolution of $N(t)$*

Fig. 2 displays a logarithmic plot of the time evolution of $1 - N(t)/4000$, the fraction of DNA beads not yet visited by the protein, for $e_{prot}/e_{DNA} = 1$ and four values of $w$ ranging between 18 nm and 135 nm. HI were taken into account for these simulations. It is seen that $N(t)$ does not deviate from the law in Eq. (3.1) even when the fraction of visited beads becomes as large as 99%. All the simulations we ran actually display a similar agreement with Eq. (3.1). We consequently extracted a rate constant $\kappa$ from each simulation by fitting the computed evolution of $N(t)$ against Eq. (3.1). These values are reported in Table 1 in units of beads/µs. This table has 24 entries, which correspond to all possible combinations obtained with four values of $w$ (18, 32, 45 and 135 nm), three different DNA-protein interaction laws (repulsive interaction, $e_{prot}/e_{DNA}=1$, and $e_{prot}/e_{DNA}=3$), and two different ways of handling HI ("off" and "on"). As will also be the case for all subsequent tables, the first number in each entry was obtained with the $\|\mathbf{r}_{j,k}-\mathbf{r}_{prot}\|\leq\sigma$ criterion, while the number in parentheses was obtained with the $\|\mathbf{r}_{j,k}-\mathbf{r}_{prot}\|\leq 1.5\sigma$ criterion. It is seen that the values of $\kappa$ vary over more than two orders of magnitude and depend very strongly on whether HI are taken into account or not. We will come back shortly to this latter point.

B - *Estimation of $\rho_{1D}$*

Klenin *et al*'s formula in Eqs. (1.3) and (1.4) depends on $\tau_{1D}$ and $\tau_{3D}$, the average times the protein spends in the bound and free states, respectively. Eq. (1.4) may be rewritten in the slightly more convenient form

$$\xi = w\sqrt{\frac{1}{2\pi}\frac{D_{1D}}{D_{3D}}\frac{\rho_{1D}}{1-\rho_{1D}}} \ , \qquad (4.1)$$

where $\rho_{1D}$ denotes the fraction of time during which the protein is attached to a DNA bead, that is $\rho_{1D}=\tau_{1D}/(\tau_{1D}+\tau_{3D})$. Values of $\rho_{1D}$ are easily extracted from the simulations by checking at each time step whether the distance between the center of the protein bead and that of any DNA bead is smaller than the threshold, that is σ or 1.5σ. The obtained values of $\rho_{1D}$ are shown in Table 2. As already emphasized in [26,27], $\rho_{1D}$ increases from nearly 0 for the repulsive potential to almost 1 for large values of the protein charge. It can also be seen in Table 2 that $\rho_{1D}$ is substantially smaller when HI are taken into account than when they are not. Stated in other word, HI tend to move the protein away from the DNA. We will see that this has a marked effect on the targeting speed. At last, it can also be noticed that the values of $\rho_{1D}$ for the largest value of $w$ (145 nm) are substantially smaller than for the three other values of $w$ (18, 32, and 45 nm), which reflects the fact that DNA segments are more widely separated and the protein consequently spends more time diffusing freely in the buffer.

### C - *Estimation of $D_{1D}$*

$D_{1D}$ is the 1D diffusion coefficient of the protein when it slides along DNA. There are obviously no such sliding events when the interaction potential between DNA and the protein is assumed to be repulsive. In this case, the protein bead may collide with, but not slide along DNA. In contrast, for the interaction potentials corresponding to $e_{prot}/e_{DNA}=1$ and $e_{prot}/e_{DNA}=3$, the protein spends a sizeable amount of time sliding along the DNA. In order to estimate $D_{1D}$, we extracted from the simulations all the sliding events with the following properties: (i) each event lasted more than 1 µs, (ii) during this time, the protein did not separate from the DNA segment by more than $\sigma$ (or $1.5\sigma$) during more than 0.07 µs, (iii) the protein bead did not reach one of the extremities of the DNA segment. We then computed the average value of $N(t)$ during these events and drew log-log plots of the time evolution of $N(t)$. A few representative plots are shown in Fig. 3. We observed that all the plots are approximately linear in log-log scales, which means that $N(t)$ evolves according to a power law $N(t) \approx \alpha t^{\beta}$. We found that $\beta$ is close to 0.5 for $e_{prot}/e_{DNA}=1$ and HI switched "on", to 0.45 for $e_{\text{prot}}/e_{\text{DNA}}=1$ and HI switched "off", to 0.40 for $e_{\text{prot}}/e_{\text{DNA}}=3$ and HI switched "on", and to 0.20 for $e_{\text{prot}}/e_{\text{DNA}}=3$ and HI switched "off". This indicates that the sliding motion is diffusive in the first case, slightly subdiffusive in the second and third cases, and very subdiffusive in the last case. This is probably connected to the fact that, when going from the first to the fourth case, the protein bead actually spends more and more time attached to the same DNA bead without moving : large average waiting times between random-walk steps are indeed sufficient to induce subdiffusion (see, for example, [46]). Except for the last scase, the time evolution of $N(t)$ can therefore be fitted with a square-root law $N(t) \approx \alpha \sqrt{t}$. The diffusion coefficient $D_{1D}$ is finally deduced from this adjusted value of $\alpha$ by using the expression of the volume of the Wiener sausage in 1D [43-45]

$$\ell(t) \approx \sqrt{\frac{16}{\pi} D_{1D}\, t}\ , \qquad (4.2)$$

where $\ell(t)$ is the length of the DNA sequence visited by the protein after time $t$. One therefore obtains

$$D_{1D} = \frac{\pi}{16} l_0^2 \alpha^2\ . \qquad (4.3)$$

Values of $D_{1D}$ obtained with this method are reported in Table 3 in units of $10^{-10}$ m$^2$ s$^{-1}$. As could reasonably be expected, the estimated values of $D_{1D}$ do not depend on the value of $w$. In contrast, $D_{1D}$ appears to be about twice larger when HI are taken into account than when they are not. Not surprisingly, $D_{1D}$ also depends to some extent on the shape and depth of the interaction potential : values of $D_{1D}$ for $e_{\text{prot}}/e_{\text{DNA}}=3$ appear to be about 40% larger than the corresponding values for $e_{\text{prot}}/e_{\text{DNA}}=1$.

D - *Estimation of $D_{3D}$*

The 3D diffusion coefficient of the protein in the buffer can be estimated in at least three different ways, namely from Einstein's formula, from the expression of the volume of the 3D Wiener sausage, and from the mean squared displacement of the protein. Einstein's formula states that the diffusion constant of a sphere of radius $a_{prot}$ in a buffer of viscosity $\eta$ is

$$D_{3D} = \frac{k_B T}{6\pi \eta a_{prot}} . \qquad (4.4)$$

When plugging $T = 298$ K, $\eta = 0.89 \times 10^{-3}$ Pa s, and $a_{prot} = 3.5$ nm in Eq. (4.4), one gets $D_{3D} = 0.70 \times 10^{-10}$ m$^2$ s$^{-1}$. It can reasonably be expected that, for repulsive DNA/protein interactions and HI switched "off", the values of $D_{3D}$ obtained from the expression for the volume of the 3D Wiener sausage (Eq. (3.7)) should be very close to this value. Such estimates, derived from the values of $\kappa$ in the "repulsive potential" column of Table 1 and Eq. (3.7) with $\delta = \sigma$ or $\delta = 1.5\sigma$, are shown in the top half of Table 4. It can be checked that they indeed agree very closely with the result of Einstein's formula.

In contrast, when HI are taken into account, the values of $D_{3D}$ obtained from the expression of the volume of the 3D Wiener sausage in Eq. (3.7) are substantially larger than those obtained from Einstein's formula (see the bottom half of Table 4). This increase of $D_{3D}$ when HI are turned "on" was rather unexpected to us, because it has long been known that HI tend to decrease the association rate between two diffusing spheres placed at short distance [47-49]. This is due to the fact that the stochastic (thermal) motions of the two particles become highly correlated, so that their relative mobility slows down. But, on the other hand, this increase of $D_{3D}$ agrees with Kirkwood-Riseman's equation, which states that HI reduce the effective friction coefficient of long DNA chains [50]. Anyway, when estimating $D_{3D}$ in the standard way, by fitting the mean squared displacement of the protein with a linear law, that is

$$\left\langle \left\| \mathbf{r}_{\text{prot}}(t) - \mathbf{r}_{\text{prot}}(0) \right\|^2 \right\rangle = 6 D_{3D} t \, , \tag{4.5}$$

we again obtained that $D_{3D}$ is larger when HI are turned "on" than "off". For example, we checked that Eq. (4.5) leads to $D_{3D} = 0.68 \times 10^{-10}$ m$^2$ s$^{-1}$ for $w$=45 nm and HI switched "off" and to $D_{3D} = 1.60 \times 10^{-10}$ m$^2$ s$^{-1}$ for HI switched "on". The dependence of the 3D diffusion coefficient of the protein on HI is a point that certainly deserves further attention for its own.

### E - *Estimation of κ from Klenin et al's formula*

All the quantities that are necessary to estimate the rate constant *κ* from Klenin *et al*'s formula for the mean time of first arrival *τ* in Eqs. (1.3) and (4.1) and the relation between *τ* and *κ* in Eq. (3.4) are now at disposal. These values are reported in Table 5 in units of beads/µs. Since there is no sliding of the protein along the DNA for the repulsive DNA/protein interaction, $\rho_{1D}$ was set to 0 in this case in Klenin *et al*'s formula, although $\rho_{1D}$ is actually small but not zero because of collisions (see Table 2). As a consequence, the "repulsive potential" column of Table 5 is similar to that of Table 1, because this column of Table 1 is used to estimate the 3D diffusion coefficient $D_{3D}$ (Table 4) according to the expression for the volume of the 3D Wiener sausage in Eq. (3.7). Moreover, for $e_{\text{prot}} / e_{\text{DNA}} = 3$ and HI switched "off", the sliding motion of the protein along the DNA is too subdiffusive to enable an estimation of $D_{1D}$. Klenin *et al*'s formula can therefore not be used in this latter case.

**5. Discussion**

In this section, we first discuss the results obtained with the dynamical model and then the extent to which these results agree with those of the kinetic model of Klenin *et al* [38]. We postpone the discussion of the conclusions concerning real systems, which can be drawn from this comparison, to the next, conclusive section.

A - *Acceleration of targeting due to facilitated diffusion and hydrodynamic interactions*

Table 6 shows the acceleration of the protein targeting process due to facilitated diffusion. This acceleration was estimated as the ratio of a given rate constant $\kappa$ for $e_{\text{prot}}/e_{\text{DNA}} = 1$ or $e_{\text{prot}}/e_{\text{DNA}} = 3$ divided by the corresponding value of $\kappa$ for the repulsive DNA/protein interaction. Table 7 similarly shows the acceleration of the targeting process due to HI. This acceleration was estimated as the ratio of a given rate constant $\kappa$ for HI switched "on" divided by the corresponding value of $\kappa$ for HI switched "off". In both cases, the values of $\kappa$ were taken from Table 1 for the dynamical model and from Table 5 for the kinetic model.

Let us first concentrate on the results obtained with the dynamical model. For a reason, which will become clear later, we discuss in the present subsection only the results obtained with the $\sigma$ threshold. For HI switched "on", the values for the acceleration of targeting due to facilitated diffusion reported in Table 6 just match those obtained in our first study [26]. They are comprised between 1.3 and 1.7 and are quite similar for $e_{\text{prot}}/e_{\text{DNA}}=1$ and $e_{\text{prot}}/e_{\text{DNA}}=3$ (note that the acceleration becomes smaller than 1 for values of $e_{\text{prot}}/e_{\text{DNA}}$ larger than 5, see Fig. 9 of [26]). Table 6 additionally indicates that the acceleration due to facilitated diffusion only marginally depends on $w$, that is the DNA concentration, when HI are switched "on". Things are, however, quite different when HI are switched "off". In this case, the acceleration due to facilitated diffusion depends significantly on $w$. When $w$ increases from 18 nm to 135 nm, the acceleration indeed increases by a factor of almost 4 for $e_{\text{prot}}/e_{\text{DNA}}=1$, and by a factor larger than 8 for $e_{\text{prot}}/e_{\text{DNA}}=3$. Moreover, the value of the acceleration depends much more sharply on the protein charge than for HI switched "on". Indeed, in the range of values of $w$ we investigated, acceleration of targeting for $e_{\text{prot}}/e_{\text{DNA}}=1$ is larger than that for $e_{\text{prot}}/e_{\text{DNA}}=3$ by a factor which varies between 3.5 and 8. More precisely, facilitated diffusion is about 10 times *slower* than 3D diffusion for $e_{\text{prot}}/e_{\text{DNA}}=3$ and $w$=18 nm, but more than 3 times *faster* for $e_{\text{prot}}/e_{\text{DNA}}=1$ and $w$=135 nm.

The crucial role of hydrodynamics is further emphasized by the values of the acceleration of targeting due to HI reported in Table 7. It is seen that, for values of $w$ close to physiological ones (30 to 50 nm), this acceleration is close to 2 for repulsive DNA/protein interactions and to 3.5 for $e_{\text{prot}}/e_{\text{DNA}}=1$, while it is as large as 20 for $e_{\text{prot}}/e_{\text{DNA}}=3$. Examination of Tables 2 to 4 suggests that the large acceleration of targeting observed when HI are switched "on" is ascribable to two rather distinct effects. First, as already noted in the preceding section, both $D_{1D}$ and $D_{3D}$ are roughly twice larger when HI are switched "on" than when they are switched "off" (see Tables 3 and 4). This, of course, accelerates the targeting process in proportion. The second effect is that HI tend to detach the protein from the DNA sequence, as can be checked by looking at the values of $\rho_{1D}$ reported in Table 2. This considerably modifies the motion of highly charged proteins. For example, for $e_{prot}/e_{DNA} = 3$ and HI switched "off", the protein spends about 90% of the time attached to DNA for physiological values of $w$. The protein remains consequently attached for most of the time to the same portion of the DNA sequence and either does not move or performs essentially 1D search, which is quite inefficient (see Eq. (4.2)). In contrast, $\rho_{1D}$ is of the order of 66% when HI are switched "on", so that, in spite of the strong electrostatic attraction exerted by DNA, the protein spends a sizeable amount of time diffusing in 3D in the buffer. Stated in other words, the reduction of $\rho_{1D}$ caused by HI allows strongly charged proteins to search efficiently for their target, while this would be forbidden by electrostatic interactions in the absence of HI.

B - *Comparison of the dynamical and kinetic models*

Let us now examine the degree of agreement between results obtained with the dynamical and kinetic models, and let us start with the results obtained when switching HI "off". For the repulsive DNA/protein interaction potential, the corresponding columns of Tables 1 and 5 are identical. This actually just reflects the facts that the values of $\kappa$ in Table 1 were used to estimate the diffusion coefficients $D_{3D}$ reported in Table 4 and that $\rho_{1D}$ was further assumed to be zero in Eq. (4.1) for repulsive DNA/protein interactions, because in this case it is not possible to derive an estimation of $D_{1D}$ from Brownian dynamics simulations. Still, when plugging in Eq. (3.8) the value of $D_{3D}$ obtained from Einstein formula (Eq. (4.4)) instead of those reported in Table 4, one again obtains "kinetic" rate constants $\kappa$ that are in excellent agreement with "dynamical" ones.

While for repulsive DNA/protein interactions the agreement between the dynamical and kinetic models does not depend on the threshold used in Brownian dynamics simulations, this is no longer the case for the interaction potential with $e_{\text{prot}}/e_{\text{DNA}}=1$. Comparison of Tables 1 and 5 indeed indicates that the agreement is pretty good for the $\sigma$ threshold, while the values of $\kappa$ estimated from Klenin *et al*'s formula are much too small for the $1.5\sigma$ threshold. This is actually also the case for all the simulations that will be discussed in the remainder of this section. Examination of Tables 3 and 4 indicates that the values of $D_{1D}$ and $D_{3D}$ derived from Brownian dynamics simulations are not sensitive to the threshold, as one would reasonably expect. In contrast, the fraction of time $\rho_{1D}$ during which the protein is attached to the DNA sequence depends strongly on the threshold. In particular, the $1.5\sigma$ threshold leads to values of $\rho_{1D}$ that are close to 1 for most of the simulations. The point is, that the values of the rate constant $\kappa$ obtained from Klenin *et al*'s formula tend towards 0 when $\rho_{1D}$ tends towards 1. This reflects the fact that the protein motion thereby switches from facilitated diffusion, for which $N(t)$ increases linearly with time, to 1D diffusion, for which $N(t)$ increases as the square root of time. Overestimation of $\rho_{1D}$ therefore essentially results in underestimation of $\kappa$. This is very clearly what happens when the $1.5\sigma$ threshold is used in Brownian dynamics simulations. In contrast, it seems that the $\sigma$ threshold leads to values of $\rho_{1D}$ that perform a better job as input values to Klenin *et al*'s formula. Therefore, we will henceforth only consider values obtained with the $\sigma$ threshold.

The top half of the last column of Table 5 is void. This is due to the fact (already discussed in Sect. 4.B) that the 1D motion of the protein for $e_{\text{prot}}/e_{\text{DNA}}=3$ and HI switched "off" is so much subdiffusive that it is neither meaningful nor practically feasible to extract diffusion coefficients $D_{1D}$ from the simulations. As a direct consequence, it is not possible in this case to derive estimates of $\kappa$ from Klenin *et al*'s formula. We are not familiar enough with the theoretical background of Ref. [40] to determine whether this is a fundamental limitation of the kinetic model, or whether Eqs. (1.3) and (1.4) can be generalized to account for subdiffusive 1D motion of the protein.

Let us now compare results obtained with the dynamical and kinetic models when HI are switched "on". The kinetic model does not explicitly incorporate HI, which rises an interesting question : are HI reducible to their effect on $D_{1D}$, $D_{3D}$, and $\rho_{1D}$ ? Stated in other words, is it sufficient to plug in Klenin *et al*'s expression the values of $D_{1D}$, $D_{3D}$, and $\rho_{1D}$ deduced from simulations with HI switched "on" to get reasonable estimates of $\kappa$ ? Comparison of the bottom halves of Tables 1 and 5 suggests that this is indeed the case. Even if the values of $\kappa$ differ in one case by a factor of 2, the agreement is generally correct.

## 6. Conclusion

In this work, we have thus shown that the dynamical model we proposed [26,27] and the kinetic model of Klenin *et al* [38] support each other, in the sense that the rate constants $\kappa$ obtained (i) directly from the simulations, and (ii) from Klenin *et al*'s formula using values of $D_{1D}$, $D_{3D}$, and $\rho_{1D}$ extracted from the simulations, are in good agreement. In particular, both models suggest that the acceleration of targeting due to facilitated diffusion is not very large for the system we considered. Table 6 indeed shows that the dynamical and kinetic models agree in predicting an acceleration comprised between 20% and 70% for physiological values of *w*, HI switched "on", and protein charges ranging from $e_{prot}/e_{DNA}=1$ to $e_{prot}/e_{DNA}=3$.

Quite obviously, one must at this point wonder how this result transfers to real DNA and proteins. The essential point is that the dynamical system corresponds to a ratio $D_{1D}/D_{3D}$ of the order of unity (see Tables 3 and 4), as is customary for translational diffusion. In contrast, the ratio $D_{1D}/D_{3D}$ for real DNA/protein systems (measured essentially by single molecule experiments) is rather of the order of $\approx 10^{-3}$ [13,14,16,18-21,51-53]. This three orders of magnitude difference may be due to the fact that in real systems the protein has to follow an helical track along the DNA, which considerably enhances the translational friction coefficient [15,54,55]. Using Klenin *et al*'s formula, acceleration of targeting due to facilitated diffusion can be written in the form

$$\frac{\kappa}{4\pi D_{3D} a c} = \frac{1-\rho_{1D}}{\frac{\pi}{2}\frac{a}{\xi}[1-\frac{2}{\pi}\arctan(\frac{a}{\xi})]} \quad , \tag{6.1}$$

where $\xi$ is given in Eq. (4.1) and $a$ is taken here as the sum of the protein and DNA hydrodynamic radii, $\sigma$. As a consequence, for a given DNA concentration (and therefore a given value of $w$), the acceleration due to facilitated diffusion depends uniquely on $\rho_{1D}$ and the ratio $D_{1D}/D_{3D}$. For each value of $D_{1D}/D_{3D}$, one can therefore search for the value of $\rho_{1D}$ for which this acceleration is maximum. The result is plotted in Fig. 4 for three different values of $w$ (18, 45 and 135 nm). The top plots show the largest acceleration of targeting (relative to 3D diffusion) that can be attained for each value of $D_{1D}/D_{3D}$, and the bottom plots the value of $\rho_{1D}$ at which this maximum is attained. It is seen that, for physiological values of $w$ (30-50 nm), facilitated diffusion cannot be faster than 3D diffusion for values of $D_{1D}/D_{3D}$ smaller than about 0.3 : maximum acceleration is indeed 1 at $\rho_{1D}=0$. For values of $D_{1D}/D_{3D}$ larger than this threshold, the maximum acceleration instead increases approximately as the square root of $D_{1D}/D_{3D}$. This maximum acceleration is furthermore attained for values of $\rho_{1D}$ close to 1/2 when $D_{1D}/D_{3D}$ is larger than about 1. At last, the maximum acceleration due to facilitated diffusion increases slowly with *w*.

For values of $D_{1D}/D_{3D}$ close to 1.5, as in our simulations (see Tables 3 and 4), Fig. 4 indicates that maximum acceleration due to facilitated diffusion is of the order of 2 for physiological values of *w*, which is exactly what we obtained (see Table 6). In contrast, realistic values of $D_{1D}/D_{3D}$ are much smaller than the 0.3 threshold, which implies, as already stated, that facilitated diffusion is necessarily slower than 3D diffusion. For such small values of $D_{1D}/D_{3D}$, Eq. (6.1) actually reduces to

$$\frac{\kappa}{4\pi D_{3D} a c} \approx 1 - \rho_{1D} \ . \tag{6.2}$$

This conclusion agrees with experimental results, which indicate that the measured apparent diffusion coefficient of molecules that do not interact with chromatin or nuclear structures (like the green fluorescent protein or dextrans) range between $10^{-11}$ and $10^{-10}$ m$^2$ s$^{-1}$ [51-53], depending on their size, as predicted by Einstein's formula, while that of biologically active molecules is instead usually reduced by a factor of 10-100 compared to this formula [56-60].

In conclusion, we have shown that even in the favorable case studied here ($D_{1D}/D_{3D} \approx 1$), the dynamical and kinetic models agree in predicting that facilitated diffusion cannot accelerate targeting by a factor larger than a few units. Extrapolation of these results to realistic values of the ratio of the diffusion coefficients ($D_{1D}/D_{3D} \approx 10^{-3}$) further indicates that in real life facilitated diffusion most certainly significantly *slows down* the targeting process. We are of course aware of the shortcomings of our dynamical model, which is a coarse-grained model with only rather simplified and non-specific interactions between DNA and the protein. Moreover, hydrodynamic interactions are also handled in a simplified way, since the complications that arise in real systems, like the fluidity of hydration water layers [61,62] and short-range lubrication effects [63], are disregarded. Still, we cannot think of physical effects that would invalidate the essential conclusions drawn in this work. We consequently agree with Prof. S.E. Halford, that it would now be time to put "*an end to 40 years of mistakes in DNA-protein association kinetics*".

**APPENDIX A : THE DEBYE-SMOLUCHOWSKI'S RATE**

If the salinity of the buffer is very low, it can be considered at first order that the electrostatic interaction between DNA and the protein is an unscreened Coulomb potential $q_A q_B / (4\pi\varepsilon r)$, where $q_A$ and $q_B$ are the charges on interacting particles A and B, and $r$ is the distance between them. Debye [64] showed that, in this case, the association rate is given by Eq. (1.2), where

$$f_{\text{elec}} = \frac{x}{e^x - 1} \tag{A.1}$$

and

$$x = \frac{q_A q_B}{4\pi\varepsilon (r_A + r_B) k_B T} . \tag{A.2}$$

By plugging $q_A = -5\bar{e}$ (the DNA electrostatic charge for about 7 bps), $q_B = 10\bar{e}$ (the typical value for a protein effective charge [65,66]), $r_A + r_B = 0.5$ nm, and $\varepsilon = 80\varepsilon_0$ in Eq. (A.2), one obtains $f_{\text{elec}} \approx 70$, which is of the same order of magnitude as the decrease in the association rate constant that Riggs *et al* measured when increasing the salinity of the buffer up to physiological values [2].

As far as we know, there exists such an explicit formula as Eq. (A.1) neither for the screened Debye-Hückel potential, nor for the sum of a screened Debye-Hückel potential and an excluded volume term, as we used in our simulations [26]. It was however checked numerically that the association rate for a screened Debye-Hückel potential is comprised between Schmolukowski's rate and Debye's one [67].

**TABLE 1 :** Values of the rate constant $\kappa$ (expressed in units of beads/µs), obtained by fitting the time evolution of $N(t)$ against Eq. (3.1), for different values of $w$, different DNA-protein interaction laws, and hydrodynamic interactions switched either "off" or "on".

| HI | $w$ (nm) | $\kappa$ (units of beads/µs) | | |
|---|---|---|---|---|
| | | repulsive potential | $e_{\text{prot}}/e_{\text{DNA}}=1$ | $e_{\text{prot}}/e_{\text{DNA}}=3$ |
| off | 18 | 2.70 (3.86) | 2.32 (2.69) | 0.30 (0.34) |
| | 32 | 0.98 (1.44) | 0.84 (0.91) | 0.121 (0.127) |
| | 45 | 0.47 (0.70) | 0.52 (0.55) | 0.086 (0.089) |
| | 135 | 0.050 (0.075) | 0.149 (0.153) | 0.037 (0.038) |
| on | 18 | 5.73 (8.40) | 7.82 (10.30) | 7.76 (8.96) |
| | 32 | 1.94 (3.00) | 2.90 (3.43) | 2.85 (3.10) |
| | 45 | 1.08 (1.68) | 1.83 (2.11) | 1.59 (1.70) |
| | 135 | 0.30 (0.38) | 0.49 (0.53) | 0.40 (0.41) |

The first number in each entry was obtained with the $\|\mathbf{r}_{j,k}-\mathbf{r}_{\text{prot}}\|\leq\sigma$ criterion, while the number in parentheses was obtained with the $\|\mathbf{r}_{j,k}-\mathbf{r}_{\text{prot}}\|\leq 1.5\sigma$ criterion.

**TABLE 2 :** Values of $\rho_{1D}$, the fraction of time during which the protein is attached to a DNA bead, for different values of *w*, different DNA-protein interaction laws, and hydrodynamic interactions switched either "off" or "on".

| HI | *w* (nm) | $\rho_{1D}$ | | |
|---|---|---|---|---|
| | | repulsive potential | $e_{\text{prot}}/e_{\text{DNA}} = 1$ | $e_{\text{prot}}/e_{\text{DNA}} = 3$ |
| off | 18 | 0.12 (0.43) | 0.60 (0.982) | 0.912 (1.000) |
| | 32 | 0.04 (0.16) | 0.60 (0.961) | 0.902 (1.000) |
| | 45 | 0.02 (0.09) | 0.61 (0.995) | 0.906 (1.000) |
| | 135 | < 0.01 (0.01) | 0.29 (0.44) | 0.46 (0.56) |
| on | 18 | 0.15 (0.44) | 0.32 (0.74) | 0.66 (0.985) |
| | 32 | 0.05 (0.17) | 0.23 (0.53) | 0.66 (0.979) |
| | 45 | 0.03 (0.11) | 0.20 (0.41) | 0.67 (0.986) |
| | 135 | < 0.01 (0.01) | 0.09 (0.19) | 0.56 (0.78) |

The first number in each entry was obtained with the $\|\mathbf{r}_{j,k} - \mathbf{r}_{\text{prot}}\| \leq \sigma$ criterion, while the number in parentheses was obtained with the $\|\mathbf{r}_{j,k} - \mathbf{r}_{\text{prot}}\| \leq 1.5\sigma$ criterion.

**TABLE 3 :** Values of $D_{1D}$, the diffusion coefficient of the protein along the DNA segment, expressed in units of $10^{-10}$ m$^2$ s$^{-1}$, for different values of $w$, different DNA-protein interaction laws, and hydrodynamic interactions switched either "off" or "on".

| HI | $w$ (nm) | $D_{1D}$ (units of $10^{-10}$ m$^2$ s$^{-1}$) | |
|---|---|---|---|
| | | $e_{prot}/e_{DNA}=1$ | $e_{prot}/e_{DNA}=3$ |
| off | 18 | 1.15 (1.30) | |
| | 32 | 1.18 (1.21) | |
| | 45 | 1.14 (1.20) | |
| | 135 | 1.15 (1.21) | |
| on | 18 | 1.94 (2.29) | 3.13 (3.18) |
| | 32 | 2.15 (2.71) | 2.82 (2.54) |
| | 45 | 1.93 (2.74) | 2.72 (2.62) |
| | 135 | 1.92 (2.39) | 2.45 (2.11) |

The first number in each entry was obtained with the $\left\| \mathbf{r}_{j,k} - \mathbf{r}_{prot} \right\| \leq \sigma$ criterion, while the number in parentheses was obtained with the $\left\| \mathbf{r}_{j,k} - \mathbf{r}_{prot} \right\| \leq 1.5\sigma$ criterion. 1D motion for $e_{prot}/e_{DNA}=3$ and HI switched "off" is too subdiffusive to be described by a diffusion coefficient $D_{1D}$ (see text).

**TABLE 4 :** Values of $D_{3D}$, the diffusion coefficient of the protein in the buffer, expressed in units of $10^{-10}$ m$^2$ s$^{-1}$, for different values of $w$, and hydrodynamic interactions switched either "off" or "on".

| HI  | $w$ (nm) | $D_{3D}$ (units of $10^{-10}$ m$^2$ s$^{-1}$) |
|-----|----------|-----------------------------------------------|
| off | 18       | 0.66 (0.63)                                   |
|     | 32       | 0.73 (0.72)                                   |
|     | 45       | 0.72 (0.71)                                   |
|     | 135      | 0.69 (0.69)                                   |
| on  | 18       | 1.40 (1.37)                                   |
|     | 32       | 1.45 (1.49)                                   |
|     | 45       | 1.65 (1.71)                                   |
|     | 135      | 4.11 (3.47)                                   |

The first number in each entry was obtained with the $\|\mathbf{r}_{j,k} - \mathbf{r}_{\text{prot}}\| \leq \sigma$ criterion, while the number in parentheses was obtained with the $\|\mathbf{r}_{j,k} - \mathbf{r}_{\text{prot}}\| \leq 1.5\sigma$ criterion. The values of $D_{3D}$ were obtained from the expression of the volume of the 3D Wiener sausage in Eq. (3.7) and the values of $\kappa$ reported in the "repulsive potential" column of Table 1.

**TABLE 5 :** Values of the rate constant $\kappa$ (expressed in units of beads/µs) obtained from Klenin *et al*'s formula for $\tau$ in Eqs. (1.3) and (4.1), the relation between $\tau$ and $\kappa$ in Eq. (3.4), and the values of $\rho_{1D}$, $D_{1D}$ and $D_{3D}$ in Tables 2 to 4, for different values of *w*, different DNA-protein interaction laws, and hydrodynamic interactions switched either "off" or "on".

| HI | *w* (nm) | $\kappa$ (units of beads/µs) | | |
|---|---|---|---|---|
| | | repulsive potential | $e_{\text{prot}}/e_{\text{DNA}} = 1$ | $e_{\text{prot}}/e_{\text{DNA}} = 3$ |
| off | 18 | 2.70 (3.86) | 2.16 (0.48) | |
| | 32 | 0.98 (1.44) | 1.08 (0.38) | |
| | 45 | 0.47 (0.70) | 0.70 (0.09) | |
| | 135 | 0.050 (0.075) | 0.21 (0.23) | |
| on | 18 | 5.73 (8.40) | 5.15 (3.93) | 4.43 (0.95) |
| | 32 | 1.94 (3.00) | 2.35 (2.76) | 2.25 (0.59) |
| | 45 | 1.08 (1.68) | 1.48 (2.02) | 1.45 (0.36) |
| | 135 | 0.30 (0.38) | 0.52 (0.69) | 0.80 (0.57) |

Since there is no sliding of the protein along the DNA for the repulsive DNA/protein interaction, $\rho_{1D}$ was set to 0 in this case in Klenin *et al*'s formula, although $\rho_{1D}$ is actually small but not zero (see Table 2). Moreover, for $e_{\text{prot}}/e_{\text{DNA}} = 3$ and hydrodynamic interactions switched "off", the sliding motion of the protein along the DNA is too subdiffusive to enable an estimation of $D_{1D}$. Klenin *et al*'s formula can therefore not be used in this latter case.

**TABLE 6 :** Acceleration of the protein targeting process due to facilitated diffusion, for both the dynamical and kinetic models.

| HI | $w$ (nm) | $e_{\text{prot}}/e_{\text{DNA}} = 1$ | | $e_{\text{prot}}/e_{\text{DNA}} = 3$ | |
|---|---|---|---|---|---|
| | | dynamical | kinetic | dynamical | kinetic |
| off | 18 | 0.86 (0.70) | 0.80 (0.12) | 0.11 (0.09) | |
| | 32 | 0.86 (0.63) | 1.10 (0.26) | 0.12 (0.09) | |
| | 45 | 1.10 (0.79) | 1.49 (0.13) | 0.18 (0.13) | |
| | 135 | 2.98 (2.04) | 4.20 (3.07) | 0.74 (0.51) | |
| on | 18 | 1.36 (1.23) | 0.90 (0.47) | 1.35 (1.07) | 0.77 (0.11) |
| | 32 | 1.49 (1.14) | 1.21 (0.92) | 1.47 (1.03) | 1.16 (0.20) |
| | 45 | 1.69 (1.26) | 1.37 (1.20) | 1.47 (1.01) | 1.34 (0.21) |
| | 135 | 1.63 (1.39) | 1.73 (1.82) | 1.33 (1.08) | 2.67 (1.50) |

Acceleration of targeting due to facilitated diffusion was estimated as the ratio of a given rate constant $\kappa$ for $e_{\text{prot}}/e_{\text{DNA}} = 1$ or $e_{\text{prot}}/e_{\text{DNA}} = 3$ divided by the corresponding value of $\kappa$ for the repulsive DNA/protein interaction. The values of $\kappa$ were taken from Table 1 for the dynamical model and from Table 5 for the kinetic model.

**TABLE 7 :** Acceleration of the protein targeting process due to hydrodynamic interactions (HI), for both the dynamical and kinetic models.

| $w$ (nm) | repulsive potential | | $e_{\text{prot}}/e_{\text{DNA}} = 1$ | | $e_{\text{prot}}/e_{\text{DNA}} = 3$ | |
|---|---|---|---|---|---|---|
| | dynamical | kinetic | dynamical | kinetic | dynamical | kinetic |
| 18 | 2.12 (2.18) | 2.12 (2.18) | 3.37 (3.83) | 2.38 (8.19) | 25.9 (26.4) | |
| 32 | 1.98 (2.08) | 1.98 (2.08) | 3.45 (3.76) | 2.18 (7.26) | 23.6 (24.4) | |
| 45 | 2.30 (2.40) | 2.30 (2.40) | 3.52 (3.84) | 2.11 (22.4) | 18.5 (19.1) | |
| 135 | 6.00 (5.07) | 6.00 (5.07) | 3.29 (3.46) | 2.48 (2.26) | 10.8 (10.8) | |

Acceleration of targeting due to HI was estimated as the ratio of a given rate constant $\kappa$ for HI switched "on" divided by the corresponding value of $\kappa$ for HI switched "off". In both cases, the values of $\kappa$ were taken from Table 1 for the dynamical model and from Table 5 for the kinetic model.

# FIGURE CAPTIONS

**Figure 1** : Schematic view of the model for $w$=45 nm. The cell is represented by a sphere of radius $R_0$=0.213 μm. DNA is taken in the form of $m$=80 segments, each segment consisting of $n$=50 beads separated by $l_0 = 5.0$ nm at equilibrium. Each bead represents 15 base pairs. The protein is modeled as a single bead (the red one in this figure).

**Figure 2** : Logarithmic plot of the time evolution of $1 - N(t)/4000$, the fraction of DNA beads not yet visited by the protein, for $e_{\text{prot}}/e_{\text{DNA}} = 1$ and four values of $w$ ranging between 18 nm and 135 nm. Hydrodynamic interactions are taken into account. It was furthermore considered that the protein is attached to bead $k$ of DNA segment $j$ if $\|\mathbf{r}_{j,k} - \mathbf{r}_{\text{prot}}\| \leq \sigma$. The dot-dashed straight lines, which were adjusted against the evolution of $1 - N(t)/4000$ for each value of $w$, were used to estimate the values of $\kappa$.

**Figure 3** : Log-log plots of the time evolution of the number $N(t)$ of different DNA beads visited by the protein during 1D sliding for various systems with $w$=45 nm. As indicated on the figure, two simulations were ran with $e_{\text{prot}}/e_{\text{DNA}} = 1$ and two other ones with $e_{\text{prot}}/e_{\text{DNA}} = 3$. Similarly, hydrodynamic interactions were taken into account for two of the simulations, but neglected for the two other ones. It was considered that the protein is attached to bead $k$ of DNA segment $j$ if $\|\mathbf{r}_{j,k} - \mathbf{r}_{\text{prot}}\| \leq \sigma$. For each simulation, $N(t)$ was averaged over several tens of sliding events with the following properties : (i) each sliding event lasted more than 1 μs, (ii) the protein did not separate from the DNA segment by more than $\sigma$ during more than 0.07 μs, (iii) the protein bead did not reach one of the extremities of the DNA segment.

**Figure 4** : Plot, as a function of $D_{1D}/D_{3D}$ and for three different values of $w$ (18, 45 and 135 nm), of the maximum value of $\kappa/(4\pi D_{3D}ac)$ that can be attained for values of $\rho_{1D}$ comprised between 0 and 1 (top plot), and plot of the value of $\rho_{1D}$ at which this maximum is attained (bottom plot). $\kappa/(4\pi D_{3D}ac)$ is evaluated according to Eq. (6.1). This ratio represents the maximum value of the acceleration of targeting, compared to 3D diffusion, which can be achieved thank to facilitated diffusion.

**FIGURE 1**

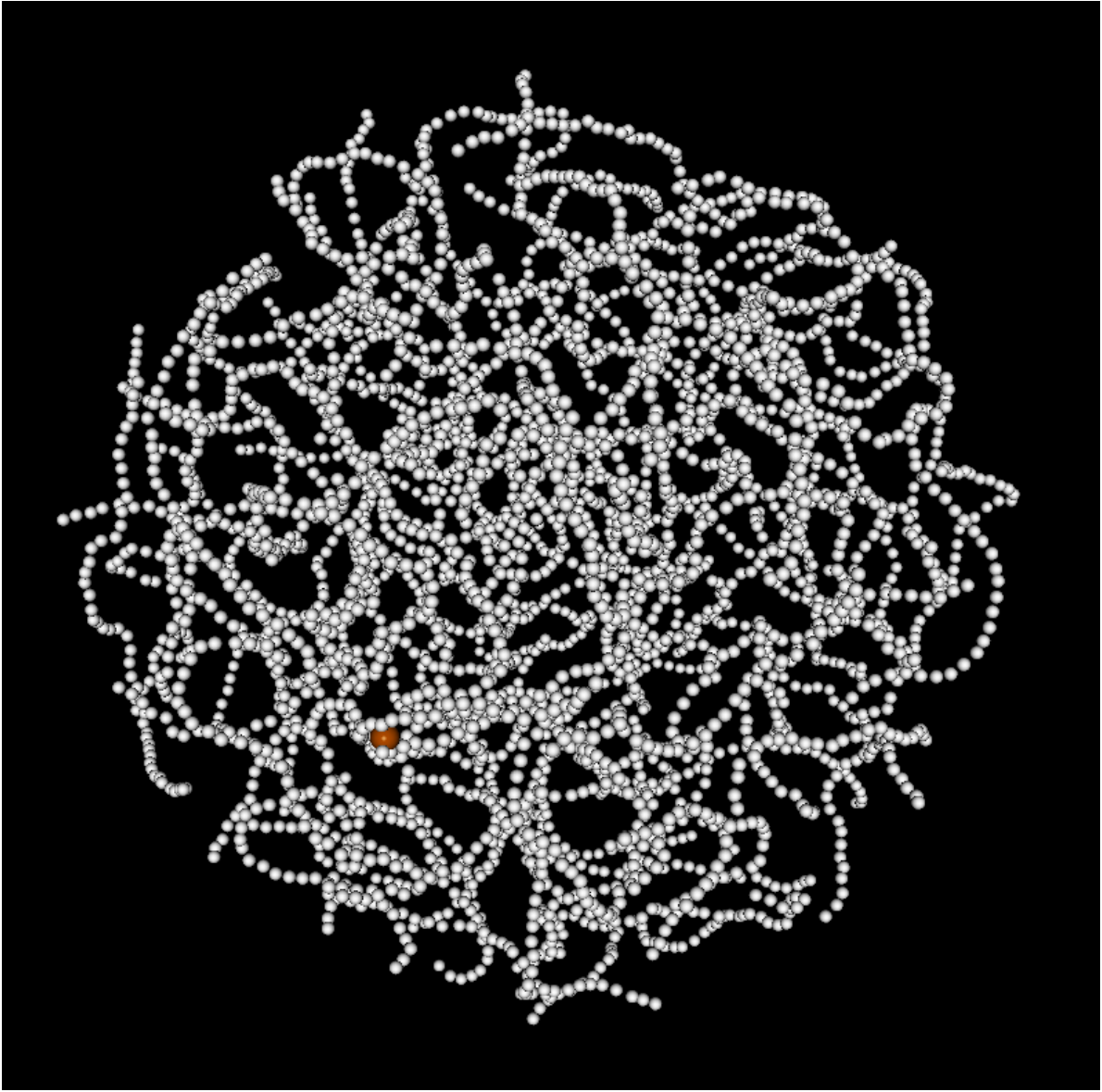

**FIGURE 2**

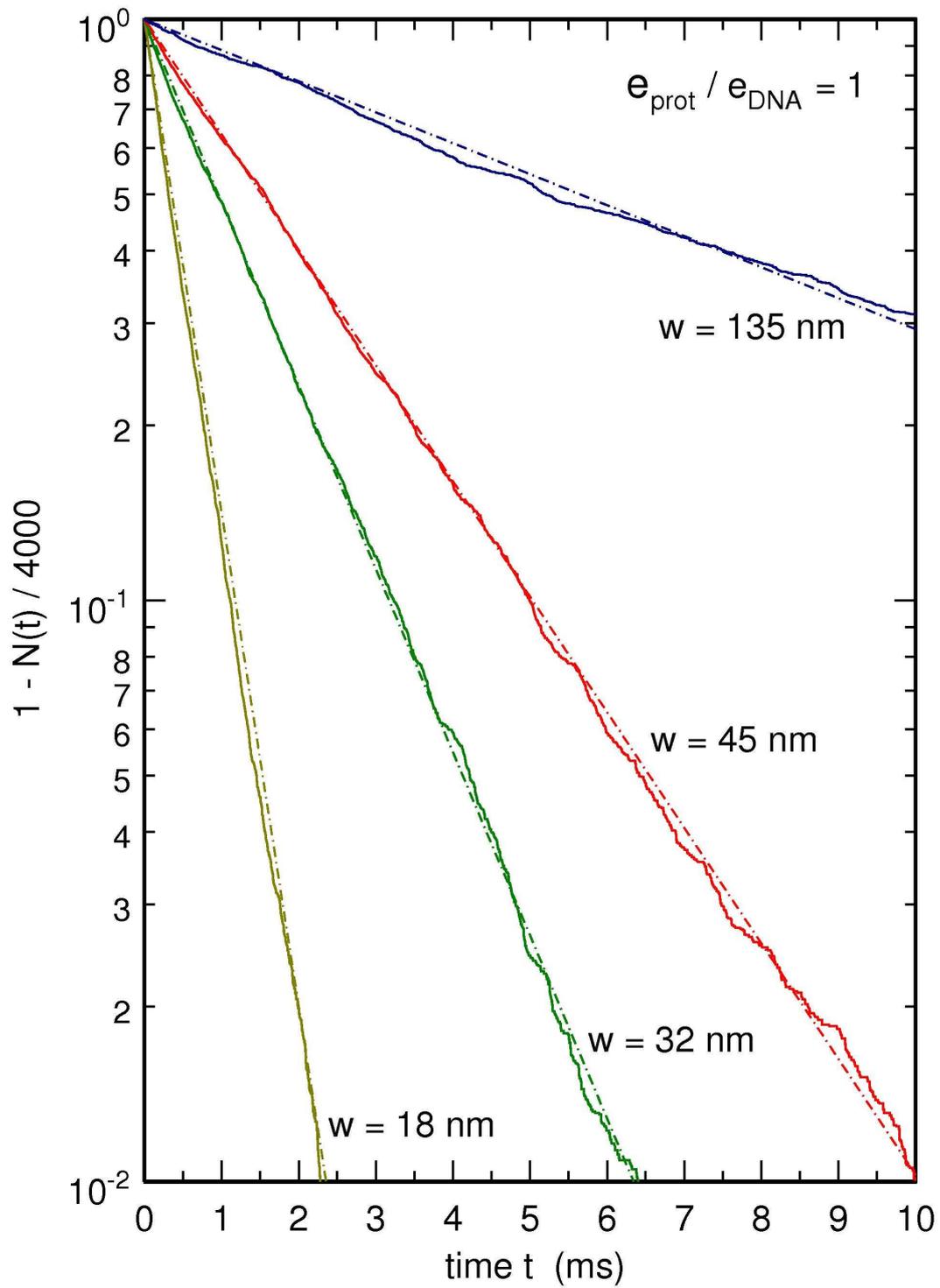

**FIGURE 3**

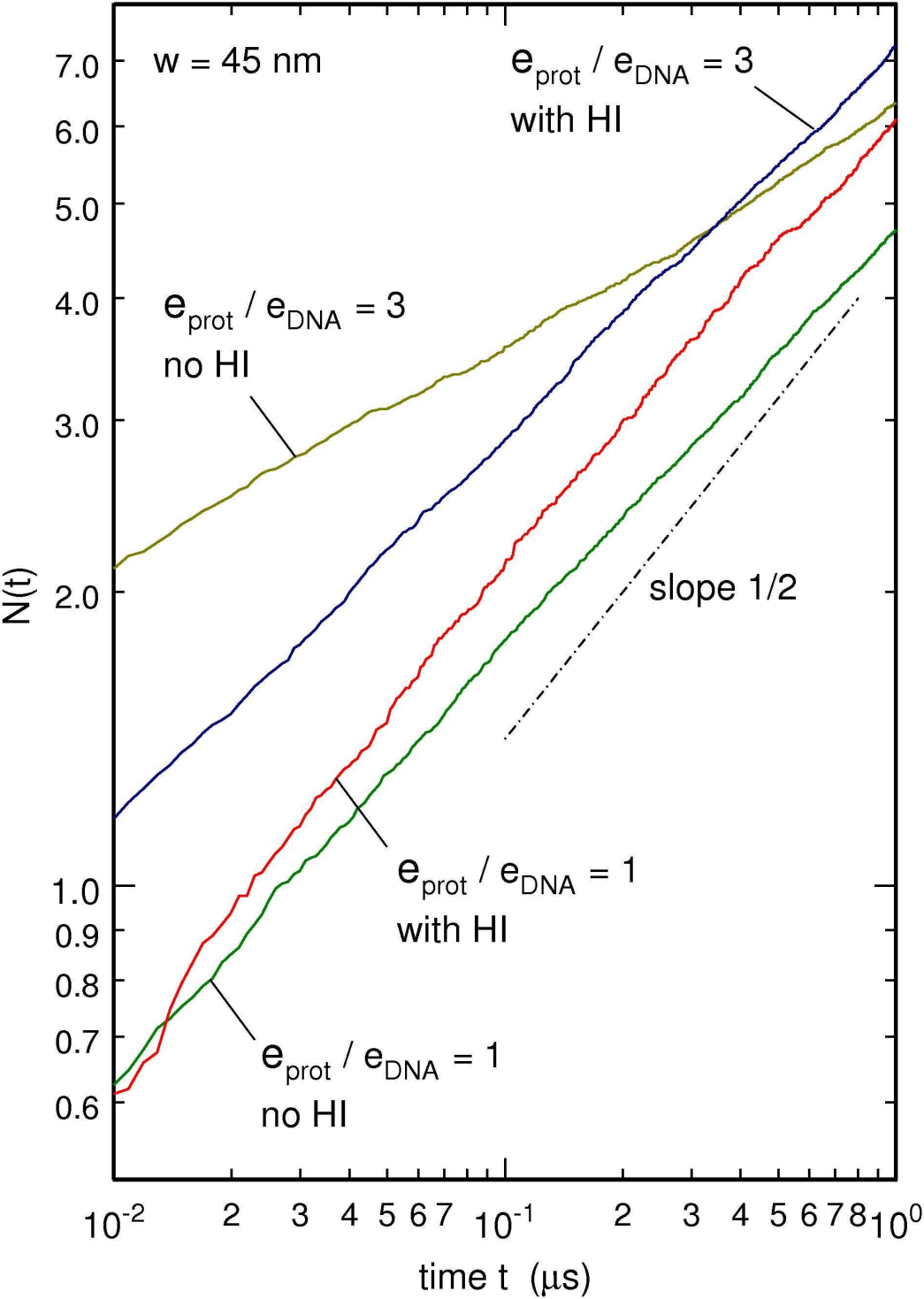

**FIGURE 4**

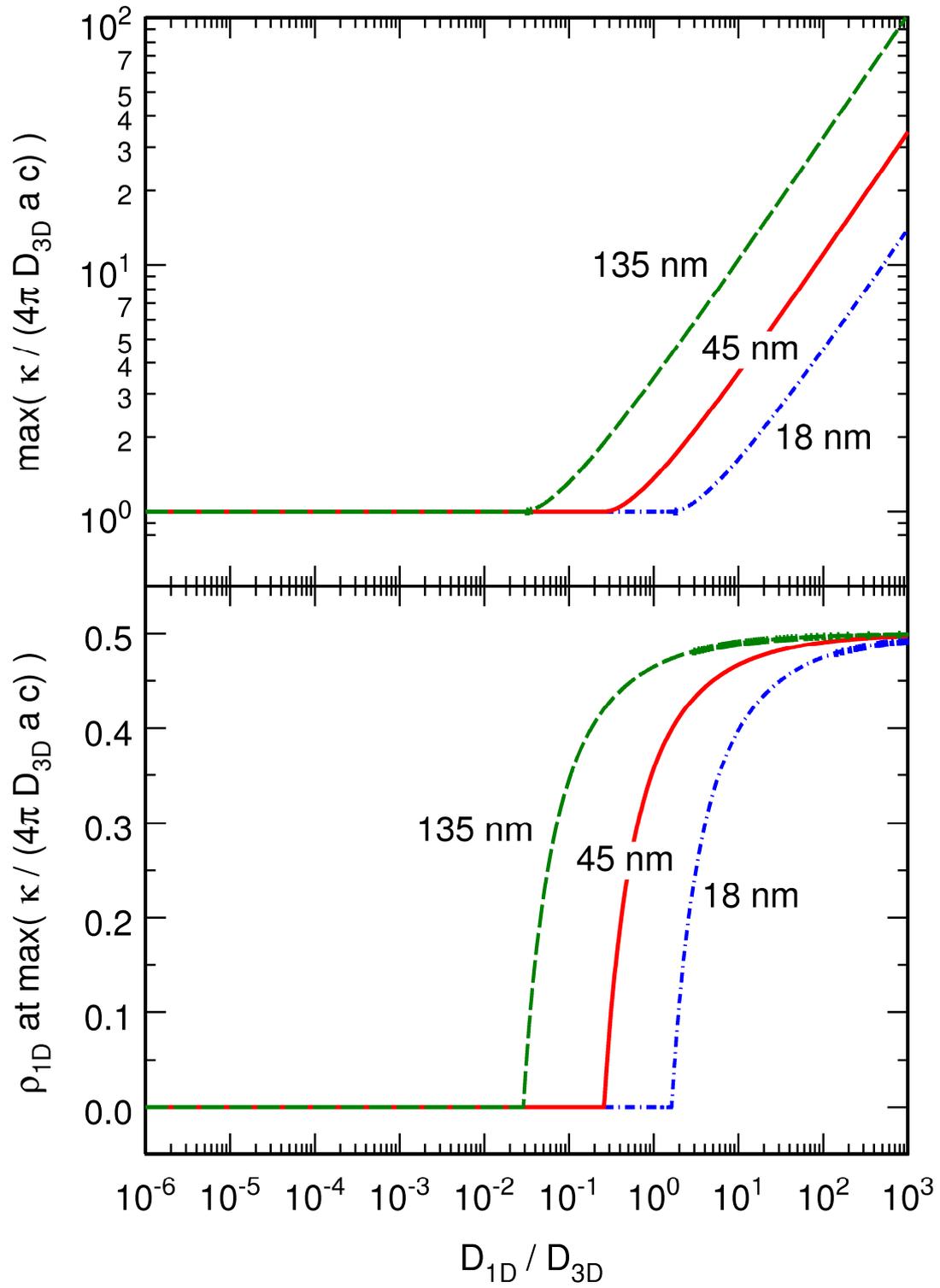